\documentclass[journal=jctcce,manuscript=article,layout=traditional]{achemso}

\usepackage[T1]{fontenc} %
\usepackage[version=3]{mhchem} %
\usepackage{xcolor}
\usepackage{hyperref}
\hypersetup{pdfborderstyle={/S/U/W 0.5}} %
\usepackage[utf8]{inputenc}
\usepackage{mathtools}
\usepackage{physics}
\usepackage{amsmath}
\usepackage{amssymb}
\usepackage{comment}
\usepackage{xcolor}

\newcommand{\upenn}{Department of Chemistry, University of Pennsylvania, 231 S. 34 Street, Cret Wing 141D, Philadelphia, Pennsylvania 19104-6243, United States}

\newcommand{\mat}[1]{\mathbf{#1}}

\author{Alec J. Coffman}
\author{Zuxin Jin}
\author{Junhan Chen}
\author{Joseph E. Subotnik}
\author{D. Vale Cofer-Shabica}
\email{valecs@sas.upenn.edu}
\affiliation[upenn] {\upenn}

\title{On the use of QM/MM Surface Hopping simulations to understand thermally-activated rare event nonadiabatic transitions in the condensed phase}

\date{February 2023}

\begin{document}

\begin{tocentry}
\centering
\includegraphics[]{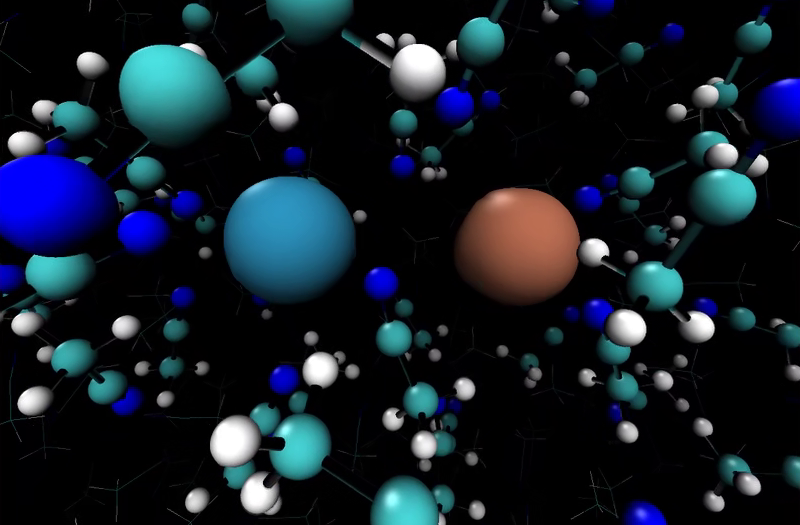}
\end{tocentry}

\maketitle

\begin{abstract}
We implement a rare-event sampling scheme for quantifying the rate of thermally-activated nonadiabatic transitions in the condensed phase.
Our QM/MM methodology uses the recently developed INAQS package to interface between an elementary electronic structure package and a popular open-source molecular dynamics software (GROMACS) to simulate an electron transfer event between two stationary ions in a solution of acetonitrile solvent molecules. 
Nonadiabatic effects are implemented through a surface hopping scheme and our simulations allow further quantitative insight into the participation ratio of solvent and the effect of ion separation distance as far as facilitating electron transfer.
We also demonstrate that the standard gas-phase approaches for treating frustrated hops and velocity reversal must be refined when working in the condensed phase with many degrees of freedom. 
The code and methodology developed here can be easily expanded upon and modified to incorporate other systems, and should provide a great deal of new insight into a wide variety of condensed phase nonadiabatic phenomena.
\end{abstract}

\section{Introduction}

A vast number of processes in chemistry occur slowly, over long timescales with large reaction barriers\cite{geissler:2002:acp}.
In order to model such processes, the standard tool is transition state theory (TST)\cite{Pechukas1981}, which offers a means to quantify rare or infrequent events using molecular dynamics; these dynamical events occur 
on time scales that would otherwise be inaccessible by integrating molecular equations of motion.
Classical transition state theory\cite{wigner1938transition} considers the ensemble average of the flux through a dividing surface at $r^{\ddagger}$ along a reaction coordinate, $r$:
\begin{equation}\label{eq:adia_TST}
    k_{TST} = \langle \dot{r} \, \theta\left( \dot{r} \right) \delta\left(r-r^{\ddagger}\right)\rangle,
\end{equation}
where $\theta(x) = 0$ for $x \le 0$ and 1 otherwise is the Heaviside function. 
Note that Eq.~\ref{eq:adia_TST} assumes
(i) that all trajectories passing through the crossing region towards products are ``reactive'',
(ii) that only a single adiabatic surface is relevant,
and (iii) that other quantum effects like tunneling are not applicable. 
Fully quantum expressions for TST are well known\cite{Miller1974,Voth1989} and in recent years, RPMD and CMD expressions for approximate quantum rate theories have been devised and implemented as well\cite{Geva2001,Craig2005,Richardson2009,Suleimanov2016,Jang2014}.

While Eq.~\ref{eq:adia_TST} is applicable to adiabatic reactions,
there are many classes of nonadiabatic problems throughout chemistry as well.
Nonadiabatic processes arise in chemical processes where simultaneous motion of electrons and nuclei cannot be separated by timescale.
Common examples of these processes are electron transfer between solvated molecules\cite{Meyer1978,barbara:1996:marcus}, electrochemical reduction at metal surfaces\cite{Hill2017}, and photochemical reactions\cite{tretiak:2014:acr}.
Such processes have broad uses in energy conversion and storage, and generalizing TST to nonadiabatic problems has been a major theme of theoretical chemistry for many years, going back to Marcus's original paper\cite{Marcus1965}. 
The basic idea is that, whereas in classical TST one can assume that the entire flux through the barrier is reactive (\latin{i.e.}\ there are no recrossings), such an assumption is obviously insufficient in the case of nonadiabatic systems.  
For nonadiabatic rare events, the process requires both $(i)$ nuclear motion over a barrier as well as $(ii)$ an electronic transition; thus, any estimate of the rate based on only a single rate-determining step will need to be suitably reduced by a nonadiabatic transmission factor, $\kappa$,
\begin{align}
\label{eq:adia_TST_kappa}
\kappa = \frac{k_{rxn}}{k_{TST}} .
\end{align}
In practice, when simulating the rate of nonadiabatic events, a suitably modified TST theory must account for the potentially large probability of recrossing, \latin{i.e.}\ the potentially small probability of a successful crossing as embodied by the probability factor $\kappa$.

There is an enormous literature on how to compute (nonadiabatic) electron transfer rates.
Note that this is a rich area of study because there are at least three different regimes, dictated by the strength of the electronic coupling and frictional effects.
$(i)$~In the low coupling regime, the dynamics are nonadiabatic in nature and strongly dictated by the electronic degrees of freedom.
$(ii)$~As the electronic coupling increases and one reaches the strong coupling regime, the process becomes more and more adiabatic.
$(iii)$~For slow solvents with large frictional effects, the dynamics can be dominated by relaxation of the nuclear coordinates even in the low coupling regime; this is the so-called Zusman limit.\cite{Zusman1980}.

Analytically interpolating between $(i)$ and $(ii)$ can be achieved  using a one-dimensional Landau-Zener transmission factor\cite{Dogonadze1971,Newton1984,Newton1991,Newton2007,Newton2008}, $\kappa_{LZ}$:
\begin{align}\label{eq:k_LZ}
P_{LZ} &= 1 - e^\frac{-2\pi \abs{V}^{2}}{\abs{v_{B}} \hbar} \\
\kappa_{LZ} &= \frac{2P_{LZ}}{1 + P_{LZ}}
\end{align}
Here, $P_{LZ}$ is the probability of hopping between diabatic states with coupling $V$ passing through a dividing surface with velocity $v_{B}$ along a reaction coordinate $\zeta$.
The factor $2P_{LZ}/(1+P_{LZ})$ arises from summing the geometric series of possible hops on the upper adiabatic surface.
Analytical interpolation within the low-coupling regimes, $(i)$ and $(iii)$, has been achieved by the theories of Zusman, Rips and Jortner\cite{Rips1987}, and Straub and Berne\cite{Straub1987}, and many others\cite{Morillo1988}.  
Notably, Ref.~\citenum{Straub1987} effectively interpolates between all three regimes.

In practice, several assumptions are usually unavoidable for the analytical transition state theories above.  
First, the models above are usually premised on the existence of well-defined diabatic states with corresponding harmonic free energy surfaces. 
For example, in the \ce{Fe^{+2}/\,Fe^{+3}} electron transfer problem, Kuharski \latin{et al.} showed that harmonic free energy surfaces could be constructed using a Hamiltonian built on pseudopotentials\cite{Kuharski1988}. 
That being said, for a model with intramolecular nuclear motion and reorganization---and which is not dominated exclusively by solvent reorganization---the assumption of harmonic free energy surfaces is questionable.
Second, the models above usually invoke the Condon approximation, \latin{i.e.}\ they assume that the diabatic coupling is a constant.
While several works have successfully explored non-Condon effects for certain models\cite{Jang2005,stuchebrukhov:1997:noncondon}, in general the presence of non-Condon effects makes the application of analytical theory far more difficult, as it is not straightforward to average over reactive trajectories experiencing very different electronic dynamics at different points in phase space. 
Roughly speaking, the assumptions of harmonic free energy surfaces and constant diabatic couplings are most appropriate for systems that can be modeled with a one-dimensional reaction coordinate.  

For complex nonadiabatic problems that do not satisfy the assumptions above, there are no tidy analytical theories and one simply needs to run dynamics near the crossing and evaluate the full rate expression in Eq.~\ref{eq:adia_TST_kappa}.
The goal of this article is to present a fully-developed computational approach for calculating such nonadiabatic rates with modern software packages as applicable to QM/MM simulations, where a quantum mechanical system can be strongly interacting with an atomistic classical environment.

An outline of this article is as follows. 
In Section~\ref{sec:nart-theory}, we review the tools and theory needed to extract nonadiabatic rates using nonadiabatic dynamical methods.  
In Section~\ref{sec:computation}, we present the software and numerical algorithms that we use to evaluate the quantities derived in the  previous section.
In Section~\ref{sec:results}, we summarize our results: we compare our rates as obtained from these simulations versus nonadiabatic rate theory and we discuss new insight into the subtleties of surface hopping with velocity reversal in the condensed phase.
Finally, in Section~\ref{sec:discussion} we conclude and discuss new avenues for future problems that can utilize the simulation framework presented within this paper.

\section{Nonadiabatic Transition State Theory (NA-TST) in Practice}\label{sec:nart-theory}
When modeling nonadiabatic rare events, one needs both an appropriate reaction coordinate and a way to model nonadiabatic dynamics for both weak and strong coupling regimes.  
We now follow what is common in the literature as far as solving the two problems above. 
The discussion below will be in the context of the simplest possible, tunable model Hamiltonian (see Eq.~\ref{eq:reac_coord}); specifically, two ions (acting as donor and acceptor) solvated by acetonitrile, with one free electron.

\subsection{Choice of reaction coordinate}\label{sec:rxncoord}
The choice of reaction coordinate is qualitatively different for nonadiabatic problems (\latin{e.g.} electron transfer) than it is for adiabatic (single-surface) problems.
For adiabatic problems, one can choose any reaction coordinate that interpolates between reactants and products, and the dividing surface is then usually chosen at the barrier between the two wells.
If the barrier is large enough and recrossings are rare enough, TST should hold well; variationally optimizing the choice of dividing surface along a reaction coordinate is also well known \cite{Pollak1990,Banushkina2015,Birkholz2015}.
For nonadiabatic problems, choosing a reaction coordinate and a dividing surface with a small recrossing rate can be much more delicate. 
After all, as discussed above, a nonadiabatic reaction is characterized by two or more potential surfaces and, for electron transfer, a successful reaction must both cross a nuclear barrier and successfully transfer an electron. Which one of these processes is more rare, and how should that inform our search for a relevant dividing surface with the fewest possible recrossings?

With these considerations in mind, one can construct several candidates for a reaction coordinate. 
From a computational point of view, the most convenient choice of reaction coordinate would be a purely nuclear coordinate. 
To that end, for the two-ion model problem investigated below (see Sec.~\ref{sec:computation}), we considered the average dipole moments along the axis connecting the two ions, the dipole moments normal to the center of the two ions, and the magnitude of the electric field at the plane of the two ions.
We further considered exponential weighting of all relevant quantities by distance from the ion pair. 
In all cases, however, we found that none of these purely nuclear reaction coordinate gave rise to meaningful surfaces (with low recrossing rates) for studying nonadiabatic rare events.

Given the difficulty constructing a reaction coordinate based on simple nuclear degrees of freedom (as well as the lack of generalizability of such an approach), we concluded that the best choice of reaction coordinate is Marcus' original idea of using the diabatic energy gap (\latin{i.e.} an electronic reaction coordinate).
To that end, given a set of two nonorthogonal atomic orbitals $(\ket{\tilde{\chi}_L}, \ket{\tilde{\chi}_R}$, we define a diabatic basis by a L\"{o}wdin orthogonalization\cite{lowdin1950orth}:
\begin{equation}
    \begin{pmatrix}
        \ket{\chi_L} \\
        \ket{\chi_R}
    \end{pmatrix}
    =
    {\begin{pmatrix}
        1 & \braket{\tilde{\chi}_L}{\tilde{\chi}_R} \\
        \braket{\tilde{\chi}_R}{\tilde{\chi}_L} & 1
    \end{pmatrix}}^{-1/2}
    \begin{pmatrix}
        \ket{\tilde{\chi}_L} \\
        \ket{\tilde{\chi}_R}
    \end{pmatrix}
\end{equation}
Our reaction coordinate below is defined as the difference in the diagonal elements of the Fock matrix in this orthonormalized AO basis:
\begin{align}\label{eq:reac_coord}
\begin{split}
\left(
\begin{array}{cc}
     F_{LL} & F_{LR} \\
     F_{RL} & F_{RR}
\end{array}
\right)  \\
\zeta = F_{LL} - F_{RR} .
\end{split}
\end{align}
This reaction coordinate can easily be generalized to larger systems of interest, provided that we can construct an adiabatic-to-diabatic transformation\cite{koppel:review:conicalbook,yarokny:1996:rmp}.
A particularly interesting feature of this reaction coordinate is that its derivative 
\begin{equation}\label{eq:gd_zeta}
    \vec{G}_{\textrm{D}} = \vec{\nabla} \zeta
\end{equation}
is spanned by the nonadiabatic coupling vector and the gradient of the adiabatic energy gap (see Eqs.~\ref{eq:d_and_zeta} and~\ref{eq:GA_and_GD}).
Thus, our choice of reaction coordinate would appear ideal for theoretical investigation.
To our knowledge, very few groups (if any) have previously used a true diabatic energy gap---as found by an adiabatic-to-diabatic transformation---for nonadiabatic rare even sampling.

As a side note, we emphasize that our reaction coordinate in Eq.~\ref{eq:reac_coord} is {\em not} exactly equivalent to the classical difference in nuclear attraction energy for charge localized on the two ions, located at $r_1$ and $r_2$, and assuming classical point charges:
\begin{equation}
\label{Eq:old_cv}
\zeta \ne
\zeta_{\textrm{point-charge}}  = 
\sum_{i} Q_{i} \left[\frac{1}{\abs{\vec{r}_{1} - \vec{R}_{i}}} - \frac{1}{\abs{\vec{r}_{2} - \vec{R}_{i}}}\right]
.
\end{equation}
In Eq.~\ref{Eq:old_cv}, one sums over all point charges $Q_i$ located at positions $\vec{R}_i$. This equation works well for the symmetric \ce{Fe^{2+}/\,Fe^{3+}} problem, especially in the diabatic limit.
However, because the gradient of Eq.~\ref{Eq:old_cv} is not spanned by the the nonadiabatic coupling vector and the gradient of the adiabatic energy gap (see also Eq.~\ref{eq:d_and_zeta}), disentangling potential energy surface effects from non-Condon effects becomes impossible with Eq.~\ref{Eq:old_cv}.
For this reason, all results below utilize Eq.~\ref{eq:reac_coord} as the reaction coordinate.

\subsection{Surface hopping dynamics for NA-TST}
Once a suitable reaction coordinate has been obtained, one must perform dynamics in order to calculate a rate constant. 
Broadly speaking, in a QM/MM scheme, dynamics must be propagated separately for the nuclear and electronic degrees of freedom.
The electronic degrees of freedom are represented by a time-dependent wave function, $\psi (t) = \sum_{i} c_{i} (t) \Phi_{i} (\vec{R} (t))$, where $\Phi_{i}$ are a predetermined set of basis functions and $\vec{R} (t)$ are the nuclear coordinates.
In the adiabatic basis, the above wave function is propagated by solving for the time evolution of the basis coefficients $c_{i}$, 
\begin{equation}\label{eq:elec_amp}
i\hbar \dot{c}_{i} = \sum_{j} c_{j} [V_{ij} - i\hbar \dot{\vec{R}} \cdot \vec{d}_{ij}],
\end{equation}
\noindent where $\vec{d}_{ij}$ is the derivative coupling vector between states $i$ and $j$ and $V_{ij}=\mel{\Phi_i}{H_{elec}}{\Phi_j}$ is the matrix element of the electronic Hamiltonian $H_{elec}$.

Broadly speaking, semiclassical theories differ by how they propagate the nuclei with respect to the electronic degrees of freedom. 
While Ehrenfest dynamics calculate the nuclear force by evaluating the average force according to the electronic coefficients $\left\{c_i\right\}$, $ \vec{F}_{Ehr} = -\Tr\left[c^{\dagger} \vec{\nabla}H c \right]$, standard Ehrenfest dynamics do not obey detailed balance\cite{Parandekar2006} (though modifications of Ehrenfest dynamics do far better\cite{Miller2015}). 
For this reason, we have worked here with surface hopping dynamics\cite{barbatti2018:NAQMMMreview}, where detailed balance is nearly obeyed\cite{Parandekar2006,Schmidt2008}.  
For such dynamics, the nuclear degrees of freedom follow Newton's equations of motion, $\mathbf{M} \ddot{\vec{R}} = - \vec{\nabla} E_{\lambda}$, where $\lambda$ is the current electronic state.
A few words are now appropriate as a brief review of fewest switches surface hopping (FSSH)\cite{Tully1990}.
FSSH handles nonadiabatic transitions by incorporating the stochastic chance of ``hops'' between electronically adiabatic surfaces at each timestep.
According to Tully, the probability of hopping from adiabatic state $\lambda$ to adiabatic state $j$ within time step $\Delta t$, $g_{\lambda \rightarrow j}$, is related to the time evolution of their electronic coefficients, $c_{\lambda/j}$:
\begin{equation}
\begin{split}
g_{\lambda \rightarrow j} &= \max\left\{0,\Delta t \frac{\dot{\rho}_{jj}}{\rho_{\lambda \lambda}}\right\} \\
&= \max\left\{0, \frac{\Delta t \left[2 \hbar^{-1} \Im(c_{\lambda} c^{*}_{j} V_{j \lambda}) - 2 \Re (c_{\lambda} c^{*}_{j} \vec{R} \cdot \vec{d}_{j \lambda})\right]} {c_{\lambda}c^{*}_{\lambda}} \right\}.
\end{split}
\end{equation}
\noindent In order to maintain energy conservation, after a hop has been recommended, the momenta of the nuclei are rescaled in the direction of the derivative coupling,
{ %
  \def\dmp{\ensuremath{
    \vec{d}_{\lambda j}^{\;\intercal} \mat{M}^{-1} \vec{p}
    }}
  \def\dmd{\ensuremath{
    \vec{d}_{\lambda j}^{\;\intercal} \mat{M}^{-1} \vec{d}_{\lambda j}
    }}
\begin{equation}\label{eq:vel_rev}
  \begin{split}
    \vec{p}' &= \vec{p} + \alpha \, \vec{d}_{\lambda j}, \\
    \alpha &=  -\frac{\dmp}{\dmd} + \sigma(\dmp)
    {\left[{\left(\frac{\dmp}{\dmd}\right)}^2 - \frac{2\Delta E}{\dmd}\right]}^{1/2} ,
  \end{split}
\end{equation}
}
\noindent where the sign function $\sigma(x)=-1$ if $x < 0$ and 1 otherwise. 
If a hop is not allowed energetically, the hop is ``forbidden.''

The scheme above is fairly straightforward in theory, and many practical tools have been implemented over the last thirty years to make simulations stable and efficient. 
Below, we will follow the protocol outlined in Ref.~\citenum{Hammes-Schiffer1995}. 
For instance, unless a hop is recommended, we never evaluate the derivative coupling matrix $\vec{d}$ in Eq.~\ref{eq:vel_rev}. Rather, we use the logarithm of the overlap matrix between adiabatic states at different times, which can be related to the time derivative matrix $\left(\mat{T}\right)_{ij} = \dot{\vec{R}} \cdot \vec{d}_{ij} = \frac{1}{dt_c}\left(\mat{log}(\mat{U})\right)_{ij}$, where $dt_{c}$ is the classical timestep.

As far as implementing surface hopping within a transition state formalism to model a rare event, the first calculations (and the major details) were provided by Hammer-Schiffer and Tully (HST)\cite{Hammes-Schiffer1995}. 
Our results below will follow a modified version of the nonadiabatic TST rate theory approach from Ref.~\citenum{Hammes-Schiffer1995}. 
The basic idea of Ref.~\citenum{Hammes-Schiffer1995} is to estimate a nonadiabatic rate by $(i)$ finding trajectories near the transition state, $(ii)$ running those trajectories backwards in time with fictitious dynamics, $(iii)$ reweighting those backwards-in-time dynamics if there was a hop and $(iv)$ running trajectories forward in time. 
From the forward dynamics, one can evaluate a corrective factor that accounts for recrossings through the dividing surface and that functions as a nonadiabatic transmission factor.  
Our methodology below will follow the HST procedure above, except for the fact there will be no reweighting, step $(iii)$; in other words, we will simply run all trajectories backwards in time and assume that there will be no hopping until the forward dynamics begin.
Mathematically, we will calculate  surface hopping rates ($k_{FSSH}$) with the following equation:
\begin{equation}
\label{Eq:SH_rate}
k_{FSSH} = \frac{\sum\limits^{N_{suc}} v_{B} e^{-\beta E_{B}}}{N_{total}Z}.
\end{equation}
\noindent Here $Z$ is the partition function on the left/reactant side of the well, $v_{B}$ and $E_{B}$ are the speed along the reaction coordinate and energy when crossing the barrier respectively, and $N_{suc}$ and $N_{total}$ are the number of trajectories that successfully go to the right/product and the total number of trajectories, respectfully.
In practice, we have previously found strong numerical results from this modified HST method\cite{Jain2015a} (where we also benchmarked the  sensitivity of the method to choice of dividing surface). 
Note that an alternative to the HST approach based on a combination of FSSH and forward sampling was recently published by Gonzalez, Dellago and co-workers\cite{Reiner2022}.

\subsection{Velocity Reversal and Overestimation of Rates}
When running surface hopping, two key phenomena that go beyond Tully's original FSSH protocol must be addressed: decoherence and frustrated hops.
Decoherence is a byproduct of treating the nuclei classically, and arises when wavepackets on different potential energy surfaces bifurcate and separate in space\cite{bittner:1995:solve,shs:1999:jpca,granucci:2007:decohere,granucci:2010:decohere}, a separation that Tully did not account for. 
In practice, this problem can often be addressed by ``collapsing'' the electronic wavefunction onto the current adiabatic electronic surface, although there is no universal criteria for determining when to collapse the electronic wavefunction\cite{Schwartz1996}.  
For cases with moderate friction, we have previously argued that decoherence is essential when calculating rates\cite{landry:2012:marcus_afssh}; for instance, without decoherence, nonadiabatic rates from direct dynamics can be far too large because of the recrossing problem\cite{Jain2015}. 
Luckily, for the ion-ion system studied below, friction is relatively weak and we will show that recrossings are not very plentiful when we start at the transition state. 
According to Ref.~\citenum{Jain2015}, we might even expect to underestimate the rate by a factor of two without decoherence. 
With this caveat in mind, implementing and investigating decoherence corrections within a fully QM/MM NA-TST theory will be the subject of a subsequent publication.

Within nonadiabatic transition state theory, the second issue, velocity reversal, is actually the more pressing issue when discussing rates of reaction. 
Earlier work by Hammes-Schiffer\cite{Schwerdtfeger2014} (following up on work by Markland \latin{et al.}\cite{Kelly2013}) demonstrated that nonadiabatic rates of reaction as calculated with surface hopping could be far too large, and that velocity reversal was the likely culprit.
As discussed above (following Eq.~\ref{eq:vel_rev}), frustrated (or ``forbidden'') hops arise when a hop is recommended from a lower energy to higher energy surface, but the nuclei lack sufficient kinetic energy in the direction of the nonadiabatic coupling.
Forbidden hops are necessary to obtain detailed balance in surface hopping calculations\cite{Parandekar2005}, and as such cannot be simply neglected.  
The most common approach for treating frustrated hops was proposed by Jasper and Truhlar\cite{Jasper2001,Jasper2002,Jasper2003,Jasper2007}, whereby the momentum is reversed along the direction of the derivative coupling, $\vec{d}_{\lambda j}$.
In this approach, momentum is reversed only if projection of the derivative coupling vector onto both the momentum and force from the target adiabat, $\vec{F}_j^a$, are in opposite directions; stated mathematically: 
\begin{align}\label{eq:rev_crit}
    (\vec{F}^{a}_{j} \cdot \vec{d}_{\lambda j})(\vec{d}_{\lambda j} \cdot \vec{p}) < 0 .
\end{align}
While Eq.~\ref{eq:rev_crit} was motivated empirically from a limited set of gas phase scattering simulations, we will argue below that a slightly different reversal criterion is likely better in general (and certainly better in the condensed phase with hundreds of atoms).
This improved criterion, which utilizes velocity instead of momentum, will be discussed in detail in Sec.~\ref{sec:results}.

\section{Computational Details}\label{sec:computation}

\subsection{QM/MM, GROMACS and  INAQS}
Solvated systems are usually large and as such certain parts of the system are often treated at the classical level.
Quantum mechanics / molecular mechanics (QM/MM) methods combine electronic structure calculations for a site of interest with a classical treatment of the environment to enable computational simulations that would not otherwise be possible.
QM/MM methods can broadly be partitioned by whether they are implicit or explicit.  
Implicit QM/MM solvent models aim to treat the solvent as a homogeneous or nearly homogeneous continuum. 
Here, we focus on explicit QM/MM solvent models that include classical force fields that model the environment in atomistic detail and offer a rich description of solute-solvent interactions.
For our explicit QM/MM simulation, 
the \latin{ab initio} Hamiltonian is of the general form,
\begin{align}\label{}
H = H_{mol} + H_{solv} + H_{mol-solv},
\end{align}
where the molecular Hamiltonian, $H_{mol} = H_{elec} + T_{nuc} + V_{nuc}$, is composed of the electronic Hamiltonian plus the nuclear kinetic and potential energies, $H_{solv}$ is the classical force-field for the solvent, and $H_{mol-solv}$ is the coupling between them, which contains a nuclear-like 1-electron term that enters into the calculation of $H_{elec}$.
Typically, the most expensive step is solving the electronic structure problem for the core system, $H_{elec}$; this electronic structure problem must then be linked  with a molecular dynamics code for the solvent.

In the present study,  we have implemented such explicit QM/MM nonadiabatic dynamics using the MD code  GROMACS\cite{gromacs45} and the linking code INAQS\cite{inaqs:1}.
GROMACS is a molecular dynamics package that performs symplectic integration via either a leapfrog or velocity Verlet scheme.
GROMACS supports many classical force fields for various solvents and ions and provides for linking to electronic structure code for evaluation of the gradients and energies of a QM subsystem.
GROMACS/INAQS is a natural choice for this problem, as it already contains the framework to drive a QM/MM simulation, with no adjustments needed to accommodate the additional requirements of performing surface hopping and nonadiabatic dynamics.
In addition to internal data analysis and sampling routines, GROMACS is also compatible with many open-source and community driven plugins (such a PLUMED\cite{plumed2}; see below) that aid in the generation of systems, visualization of data, and many other features.

The code that links GROMACS and our in-house electronic structure package is INAQS (\textbf{I}nterface for \textbf{N}on\textbf{A}diabatic \textbf{Q}M/MM in \textbf{S}olvent); the details of this software were published recently in JCTC\cite{inaqs:1}.
INAQS bridges the QM and MM parts of the calculation and implements the details of surface hopping. 
In particular, INAQS propagates Eq.~\ref{eq:elec_amp},
determines whether a hop occurs at a given time step, and rescales/reverses the velocities/momenta if a hop occurs, all in a manner consistent with the GROMACS calculations.
As discussed in Ref.~\citenum{inaqs:1}, INAQS is now fully capable of running QM/MM dynamics connecting GROMACS and the electronic structure package Q-Chem, and a simulation of a larger system with an \latin{ab initio} Hamiltonian evaluated with Q-Chem will be reported in a future publication.

As proof of concept, in this paper we study a very simple system; two ions (acting as donor and acceptor) solvated by acetonitrile, with one free electron.
This system is inspired in no small part by the canonical problem of electron transfer between \ce{Fe^{+2}/\,Fe^{+3}}, as originally popularized by Marcus\cite{Marcus1956} and then many others\cite{Kuharski1988,Logan1983,Bader1989}.
Additionally, a further simplification made is that the ions will be static, \latin{i.e.} fixed in three-dimensional space.
The decision to immobilize the ions allows us to probe the contributions of thermal effects and solvation on the nonadiabatic process without regard for the motion of the donor/acceptor pair.
Acetonitrile is chosen as the solvent due to its polar, aprotic nature.
Force field parameters for acetonitrile were taken from Caleman \latin{et al.} \cite{Caleman2012a}, available at \url{http://virtualchemistry.org}.
For this problem, we generate QM gradients and energies by direct diagaonlization in an atomic basis
using an in-house electronic structure package.

\subsection{Umbrella sampling and WHAM}
When evaluating a reaction rate, the first step is often to evaluate the barrier for the reaction coordinate.
One common means of evaluating such a free energy is umbrella sampling\cite{Torrie1977,chandler:statmech}, whereby the underlying potential energy surface is replaced with a biased energy that allows one to access regions of the original energy landscape that were previously energetically inaccessible.
When a series of umbrella sampling simulations are performed over a wide range of the reaction coordinate, a weighted histogram analysis method (WHAM)\cite{Ferrenberg1989,Kumar1992} can be used to obtain an unbiased probability distribution.
This unbiased probability distribution can then be mapped to a free energy surface (FES), in this case the potential of mean force over the chosen reaction coordinate. Below, we use the FES generated by umbrella (importance) sampling and WHAM as the starting point for our NA-TST calculation.

\subsection{Simulation pipeline}
The following summarizes the steps we have taken to generate  the nonadiabatic rates for the model ion-ion system above. The step-by-step outline below should be very general and directly applicable to future  studies. As described below, the general approach is to first posit a reaction coordinate (in our case, defined by Eq.~\ref{eq:reac_coord} above), second validate that the free energy surface of that reaction coordinate has a large barrier separating reactant and product wells, and third run dynamics.
We will now outline explicitly the details of each step in this pipeline for complete repeatability.

\begin{figure}[!htb]
\includegraphics[width=16cm,height=10.67cm]{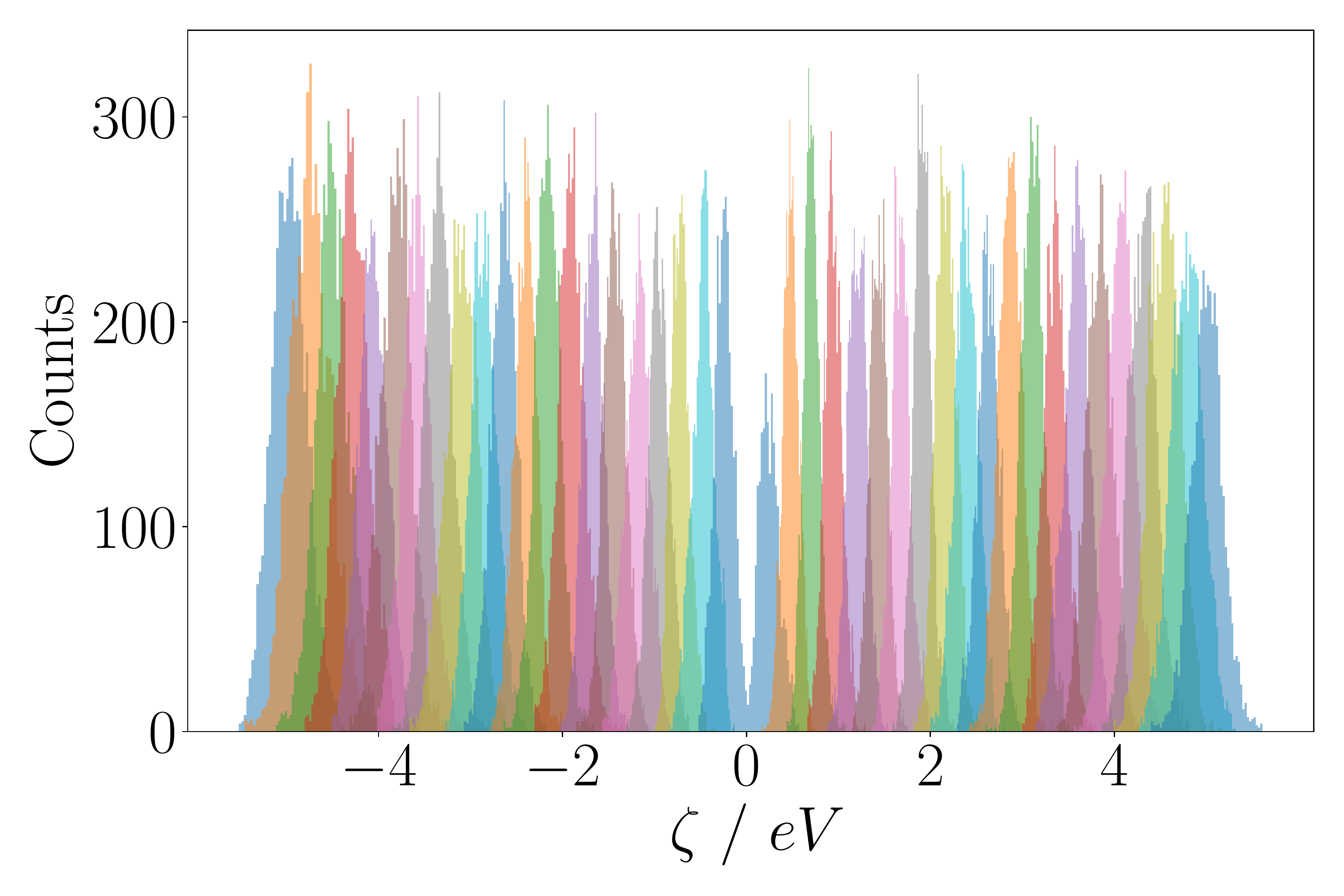}
\caption{\label{Figure:histograms_55} Umbrella sampling histograms used for generating FES for ions separated by 5.5~\AA, showing every 5th window out of all 201 windows to allow for visual clarity. 
Note that the entire 201 histograms overlap much more significantly than those shown here, a necessary requirement for successful WHAM inferences and analysis.}
\end{figure}

\begin{figure}[!htb]
\includegraphics[width=16cm,height=10.67cm]{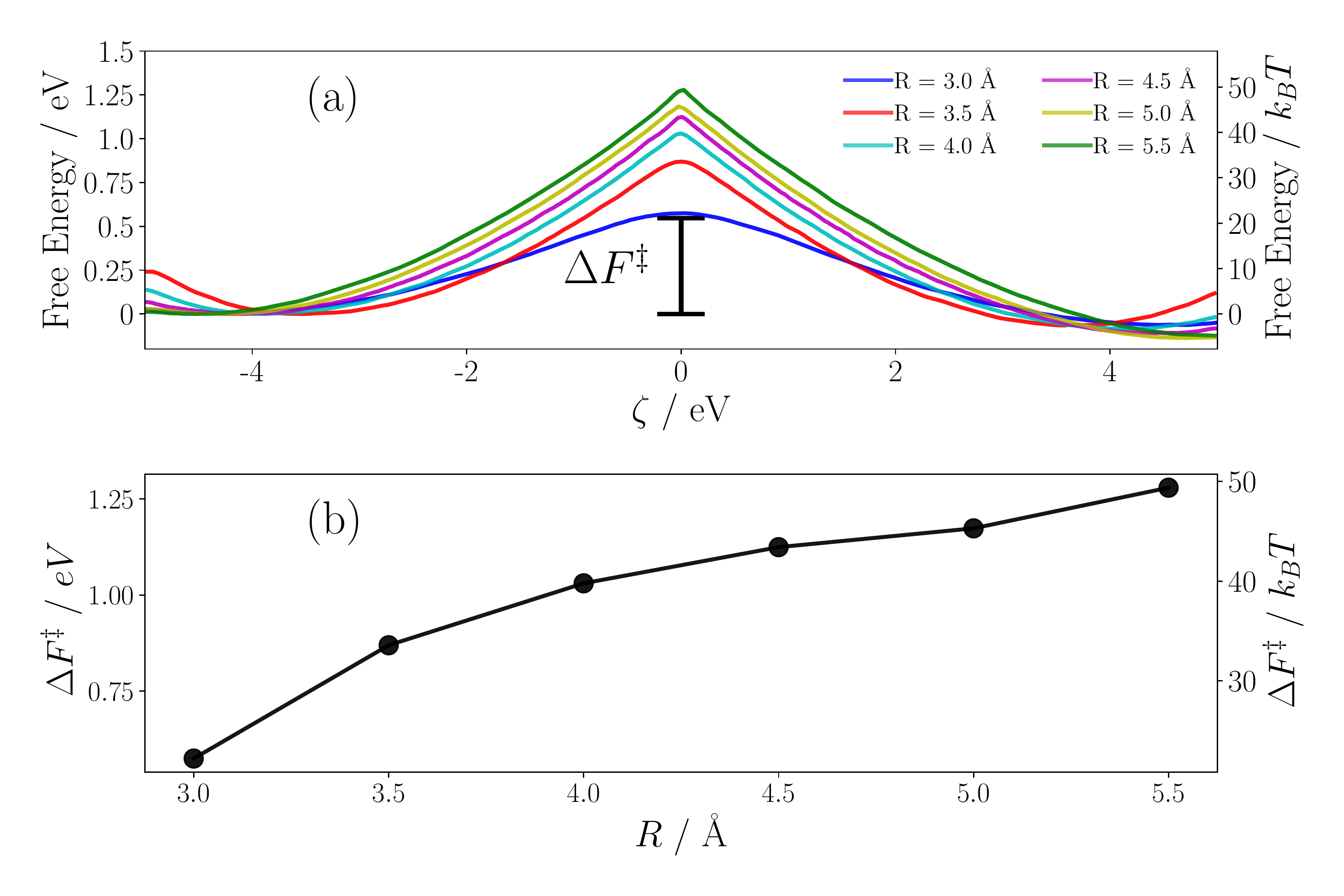}
\caption{\label{Figure:FES_all} (a) The FES for electron transfer (ET) between two fixed ions, separated by 3 to 5.5~\AA, centered in a 35~\AA \ box and solvated by acetonitrile.
The barrier height is shown here by $\Delta F^{\ddagger}$.
Note the slight asymmetry in the left and right well; this asymmetry is not a function of this specific ion ET reaction, but instead due to asymmetric boundary conditions (finite size and distribution effects of the acetonitrile solvent).
(b) The barrier heights from (a), plotted as a function of ion separation distance.
The barrier heights grow rapidly at small ion separation distances, and begin to level off at larger ion separation distances.}
\end{figure}

\begin{enumerate}
\item
First, a 35~\AA \ box of acetonitrile was generated using GROMACS, with two ions placed in the center of the box, such that the center of mass (COM) of the two ions was at exactly the center of the 35~\AA \ box.
Multiple separation distances of the two ions were examined, but the ions were always placed in the box to satisfy the condition that their COM was at the center of the simulation box.
\item
Next, the atoms of all acetonitrile molecules at a distance of greater than 12~\AA\ from the center of mass of the two ions were designated as ``frozen'' atoms.
These atoms remained stationary throughout the entire simulation pipeline, and acted as a means of keeping the remainder of the solvent molecules (designated ``mobile'' solvent) from evaporating over long timescales.
This effective droplet model\cite{Sankararamakrishnan2000} required no external restraint on the frozen solvent molecules, as long as they were packed densely enough so as not to allow the mobile solvent room to leak out of the mobile region of the simulation box. 
\item
After initialization of the system, a fully classical MM NVE simulation was run for 500~ps (using GROMACS), with an external constraint on the reaction coordinate of the system, $\zeta$, utilizing the MD plugin PLUMED\cite{Bonomi2019,Tribello2014}.
Such a constraint is necessary so as to allow sampling over a barrier and generate a FES along the reaction coordinate using umbrella sampling.
PLUMED incorporates this constraint by applying a harmonic restraint on the reaction coordinate variable $\zeta$ of the form $\frac{k}{2} (\zeta - \zeta_{c})^{2}$, where $k$ and $\zeta_{c}$ are parameters used to choose the strength and center of the restraint, respectively.
$k$ was chosen to be large enough such that there was not much spread along the reaction coordinate, but also small enough that adjacent simulations windows overlap non-trivially.
For our model problem, $k = 500 \frac{\textrm{kJ}}{\textrm{mol}\cdot \textrm{eV}^{2}} $
gave suitable distributions for use in analysis.
For each separation distance, we performed 201 simulations, centered every 0.05~eV along the reaction coordinate, from $\zeta_{c} = -5\, \textrm{eV}$ to $\zeta_{c} = 5 \, \textrm{eV}$.
For a visualization of the resultant (reasonable) distribution of WHAM histograms, see Fig.\ \ref{Figure:histograms_55}.
The FESs for ion separation distances between 3~\AA \ to 5.5~\AA \ are shown in Fig.\ \ref{Figure:FES_all}, along with the barrier heights as a function of ion separation distance.
\item
Having generated a meaningful FES, we performed dynamics and sampling along each surface.
Following the theory outlined above, simulations were initialized at the barrier of the FES.
Using PLUMED to constrain the value of the reaction coordinate, we initialized 500 trajectories at the barrier, and ran fully classical MM dynamics for 100~ps to generate statistically independent trajectories.
We then ran QM/MM trajectories, and after a 5~ps relaxation, we saved the configuration corresponding to the simulation frame nearest to the barrier top (corresponding to $\zeta = 0 \, \textrm{eV}$ with the electron nearly equally distributed between the two ions) with two constraints: $(i)$ $-.005\,\textrm{eV}< \zeta < 0 \,\textrm{eV}$, and $(ii)$ $\dot{\zeta} > 0$.
These two additional constraints guaranteed that these trajectories all began on the same side of the barrier, and were moving with velocity towards the product.
These configurations at the barrier were then taken as the starting point for SH dynamics.
\item
Before running SH trajectories, following Ref.~\citenum{Hammes-Schiffer1995}, one requires \textit{a priori} knowledge of the quantum amplitudes used in computing hop probabilities.
To that end, following Ref.~\citenum{Jain2015a}, trajectories were run backward in time along the ground state adiabat, starting at the barrier with the identity matrix for the electronic states of the system.
This matrix was propagated using the time-reversed Schr\"{o}dinger equation, and once the trajectory reached the well bottom the electronic coefficients were obtained.
If a trajectory ever crossed the barrier at any point in the back-propagation, that trajectory was removed from future analysis.

\item
Once the electronic amplitudes were obtained, trajectories were run forward in time with the ability to hop between adiabats using FSSH.
In addition to information about when and where hops occur along this reaction coordinate, we recorded many other observables (derivative coupling vectors, adiabatic and diabatic gradients, velocities, barrier recrossings, and product/reactant ratios).
\end{enumerate}

\section{Results}\label{sec:results}
500 independent trajectories were run for each ion separation distance of 3.0~\AA \ to 5.5~\AA.
We will now report our results focusing on the role of solvent, the accuracy of the resulting NA-TST rates, and the choice of velocity reversal (which emerges as a crucial feature in these surface hopping calculations).

\subsection{Participation ratio of solvent}
\begin{figure}[!htb]
\includegraphics[width=16cm,height=10.67cm]{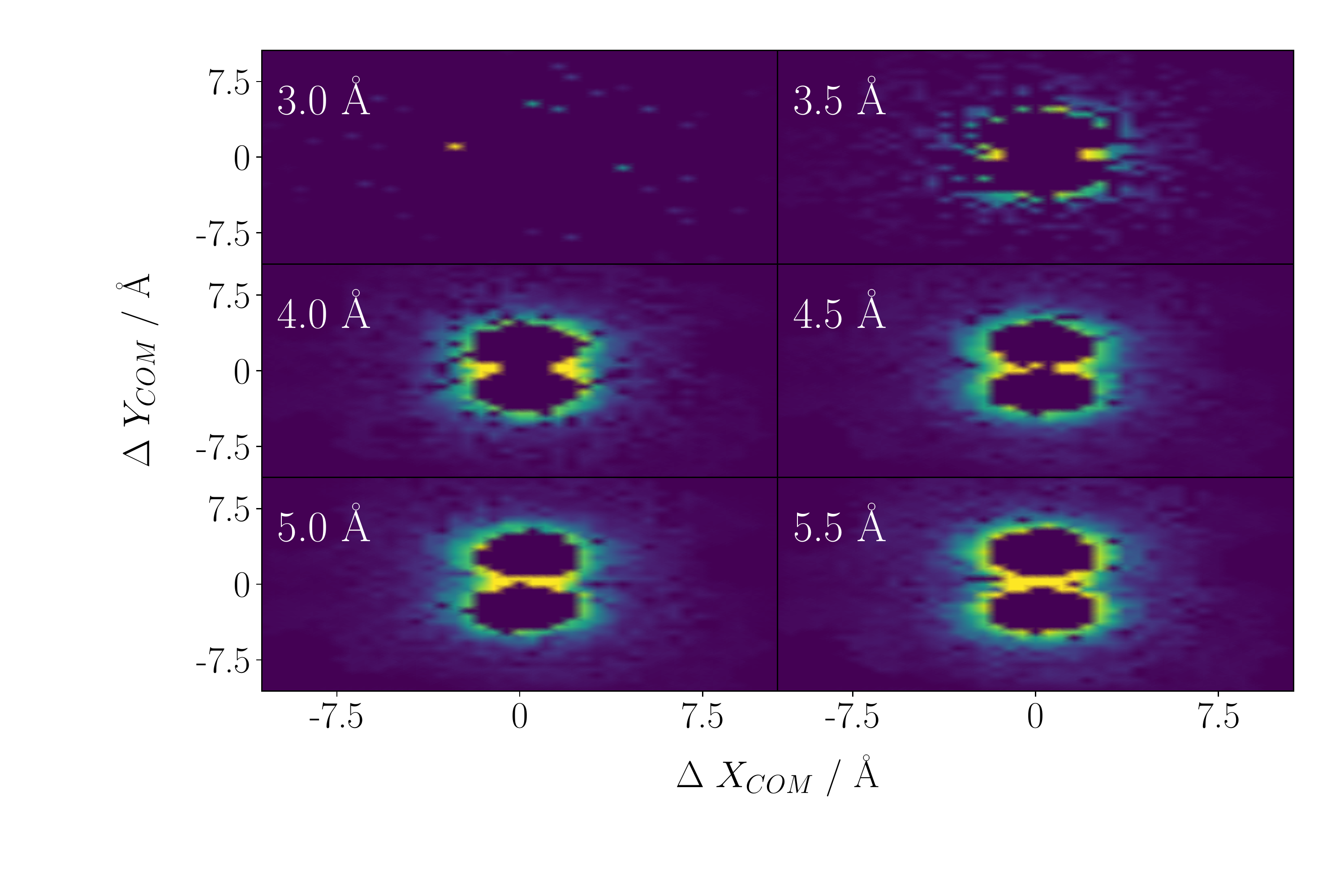}
\caption{\label{Figure:heat_map_all} A 2D cross sectional heat map (fixed at $z_{COM} = 0\,\textrm{\AA}$, the center of the simulation box) showing the relative contribution of acetonitrile molecules to each hopping event, for ion separation distances from 3.0 to 5.5~\AA.
The relative contribution here is measured by the normed contribution to the nonadiabatic coupling vector from each acetonitrile in the system, averaged over all hopping events over roughly 500 independent simulations per separation distance (see Eq.~\ref{eq:heat_map}).
These heat maps show that almost all participating solvent molecules are those nearest to the ion pair, and that at large enough separation distances, solvent molecules interstitial to the two ions can also participate.}
\end{figure}

An interesting feature that comes from performing QM/MM with explicit solvent is the ability to probe just exactly how the solvent influences hopping between adiabatic states.
One expects that, at small ion-ion distances, solvent will remain on the periphery; however, at longer ion-ion distance, solvent can intercalate.
One way to visualize this feature is to look at a spatial map of the contribution of each solvent molecule to the overall nonadiabatic coupling vector (NAC) at the time of each hopping event, $I(x,y)$. Mathematically, to define such a map, we first define the relative derivative coupling contribution for a given atom $A$, $D_{(\alpha)}^A$ at the time of a hop (indexed by $\alpha$):
\begin{equation}
\begin{split}
d_{(\alpha)}^{A} &= \sqrt{
  {\left(d_{(\alpha)}^{Ax}\right)}^2 +
  {\left(d_{(\alpha)}^{Ay}\right)}^2 +
  {\left(d_{(\alpha)}^{Az}\right)}^2
  } \\
  D_{(\alpha)}^{A} &= \frac{d_{(\alpha)}^{A}}{\sum_{A}^{N_\textrm{atoms}} d_{(\alpha)}^{A}}
\end{split}
\end{equation}
Next, we discretize space into a uniform $40 \times 40 \times 40$ grid, with cell volumes of 0.670 \AA$^{3}$.
Using a characteristic function $\gamma$ to demark the region $(x \pm \Delta, y \pm \Delta, 0 \pm \Delta)$, we define $I_{(\alpha)}(x,y)$ as roughly the contribution of the derivative coupling divided by the local density of atoms:
\begin{equation}
\begin{split}
\gamma(\vec{R}; x,y) &\equiv \left\{
  \begin{split}
  1& \quad \textrm{if} \;
  \abs{x-R_x} < \Delta \; \textrm{and} \;
  \abs{y-R_y} < \Delta \; \textrm{and}  \;
  \abs{R_z} < \Delta \\
  0& \quad \textrm{otherwise}
  \end{split}
  \right. \\
  \mathcal{N}(x,y) &\equiv \sum\limits_{\alpha} \sum\limits_{A} \gamma(\vec{R}_A,x,y) \\  
  I_{(\alpha)}(x,y) &\equiv \frac{\sum_A{\gamma(\vec{R}_A,x,y) D_{(\alpha)}^{A}}}{\mathcal{N}(x,y)}
  \end{split}
\end{equation}
The normalization factor $\mathcal{N}(x,y)$ is the total number of atoms in the grid point at $(x,y)$ summed over all hopping events $\alpha$. 
Finally, we sum over all hopping events and recover the characteristic contribution of each region to the overall nonadiabatic coupling at the time of a hop:
\begin{equation}\label{eq:heat_map}
    I(x,y) = 
  \sum\limits_{\alpha}^{N_\textrm{hops}} I_{(\alpha)}(x,y).
\end{equation}
In Fig.~\ref{Figure:heat_map_all}, we present a heat map for Eq.~\ref{eq:heat_map} showing that the nearest solvent molecules contribute most substantially to the electron-vibrational coupling.
The large circles in the middle of each heat map in Fig.~\ref{Figure:heat_map_all} are 
cavities where solvent is excluded by the Lennard-Jones potentials of each ion. 
Moreover, at large distances, it would appear that we see intercalation of solvent between the ions.

While the above visualization provides a qualitative picture of solvent mediation in this system, a more quantitative descriptor of the influence of the solvent can be found as well.
One simple approach is to calculate the participation ratio\cite{Bell1970} of the solvent with regards to a salient observable of the system.
Guided by Eq.~\ref{eq:elec_amp}, in which the term promoting population transfer between states $i$ and $j$ is $\vec{v}\cdot \vec{d}_{ij}$, we use the projection of the nonadiabatic coupling vector (NAC) onto the velocity of a given solvent atoms as a measure of its contribution to the hopping event:
\begin{equation}
\begin{split}
\label{Eq:part_rate}
P_{sol} = \frac{\sum_{j}^{N} n_{j}^{2}}{N} \\
n_{j} = \frac{\vec{d}_{j} \cdot \vec{v}_{j}}{\sum_{i}^{N} \vec{d}_{i} \cdot \vec{v}_{i}}
\end{split}
\end{equation}
\noindent Here $\vec{d}_{j}$ and $\vec{v}_{j}$ are the NAC and velocity vectors for molecule $j$, respectively, and $N$ is the total number of solvent molecules.

\begin{figure}[!htb]
\includegraphics[width=10.67cm,height=10.67cm]{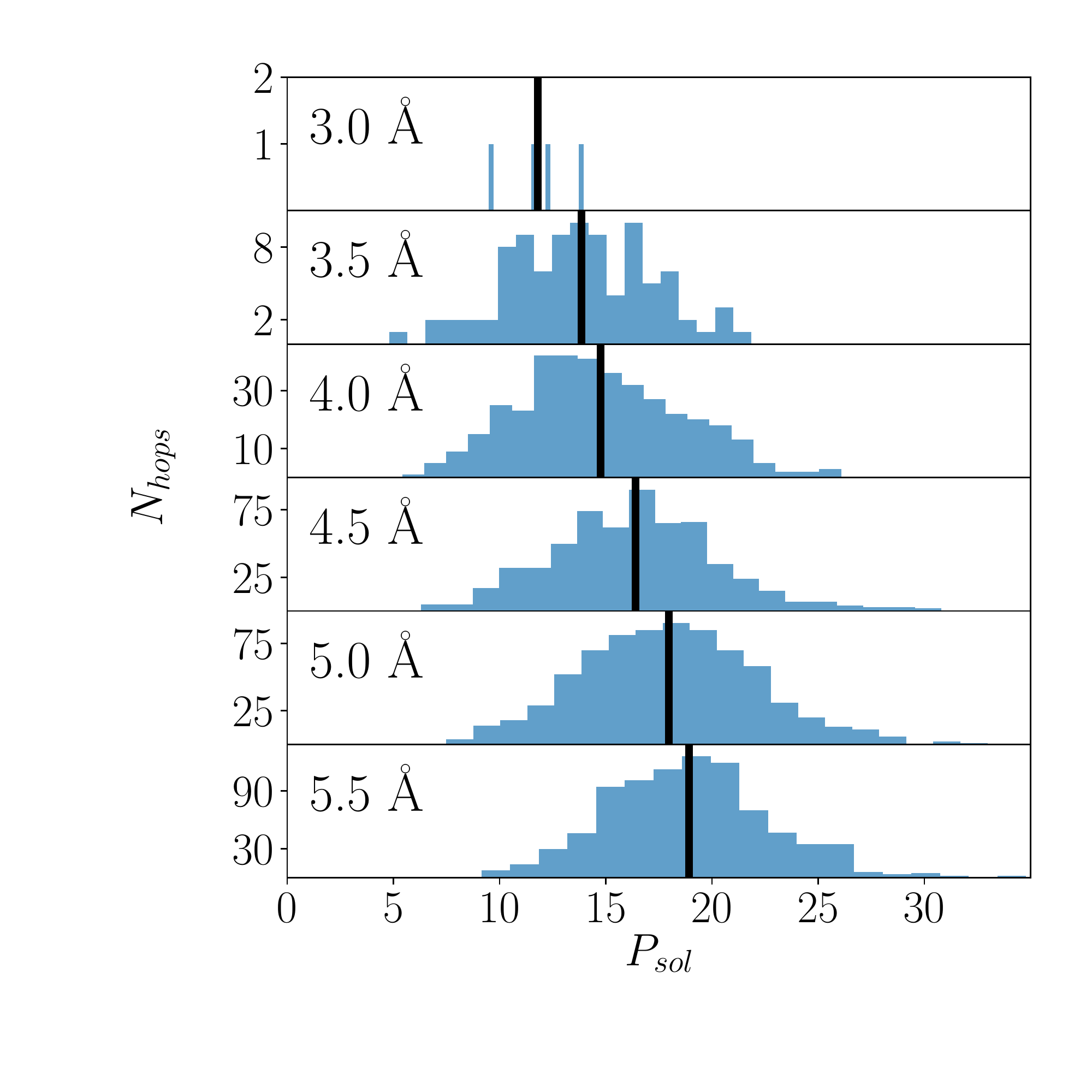}
\caption{\label{Figure:part_rate_all} Histogram of participation ratios of acetonitrile for ion separation distances from 3.0~\AA \ to 5.5~\AA.
The metric for participation ratio here is the projection of the solvent velocity vector onto the nonadiabatic coupling vector, Eq.~\ref{Eq:part_rate}.
Data are taken from hopping events with roughly 500 independent simulations per separation distance.
This figure suggests that, on average, about fifteen acetonitrile molecules contribute to the hopping between adiabatic states, although this number slightly rises slightly as a function of increasing separation distance.
The increase in participation ratio at larger ion separation distances can be attributed to the inclusion of interstitial solvent molecules, as visually illustrated in Fig.~\ref{Figure:heat_map_all}.
The sparsity of the 3.0 \AA \ histogram arises due to the fact that there are only four hopping events over the roughly 500 surface hopping simulations at that ion separation distance.}
\end{figure}

In Fig.~\ref{Figure:part_rate_all}, we plot a histogram of participation ratios ($P_{sol}$) over roughly 500 unique hopping events for simulations run with a ion separation distance of 3~\AA \ to 5.5~\AA. 
Using this metric for calculating participation ratio, we find that on average about fifteen acetonitrile molecules contribute meaningfully to the hopping between adiabatic states at small ion separation distances, and this number increases to about twenty acetonitrile molecules at larger separation distances.

Now, there is one important nuance that must be addressed. 
Above, we have quantified the participation ratios for the solvent at the moment of a hop, and these ratios give an indication of the number of solvent atoms driving the nonadiabatic transition---that is \emph{preventing} electron transfer. 
In reality, however, usually a chemist is curious to understand how to \emph{promote} the charge transfer, and instead would like to quantify (somehow) the number of solvent molecules that allow an adiabatic (rather than nonadiabtic) barrier crossing.
To that end, one can in principle compute the distribution of participation ratios at the moment of the last barrier crossing (rather than at the moment of a hop).  
We have performed such a calculation at ion separation 4.0~\AA\bibnote{Only at $R=4.0\,\textrm{\AA}$ do we have sufficient trajectories that are reactive (272) or nonreactive (194) that we can make a statistically meaningful comparison.} and we find that that the participation ratios have identical mean values of 15.8 for \emph{both} reactive and nonreactive trajectories.
In other words, within the tools available, we cannot identify  any special, set solvent configuration that promotes electron transfer; it would appear that reactive and nonreactive trajectories are distinguished more than anything else simply by the random probability for a trajectory to hop as it traverses a one-dimensional reaction barrier---though as we will discuss below, there are certainly some very multidimensional aspects to this problem. 

\subsection{Rate of barrier crossing}
\begin{figure}[!htb]
\includegraphics[width=16cm,height=10.67cm]{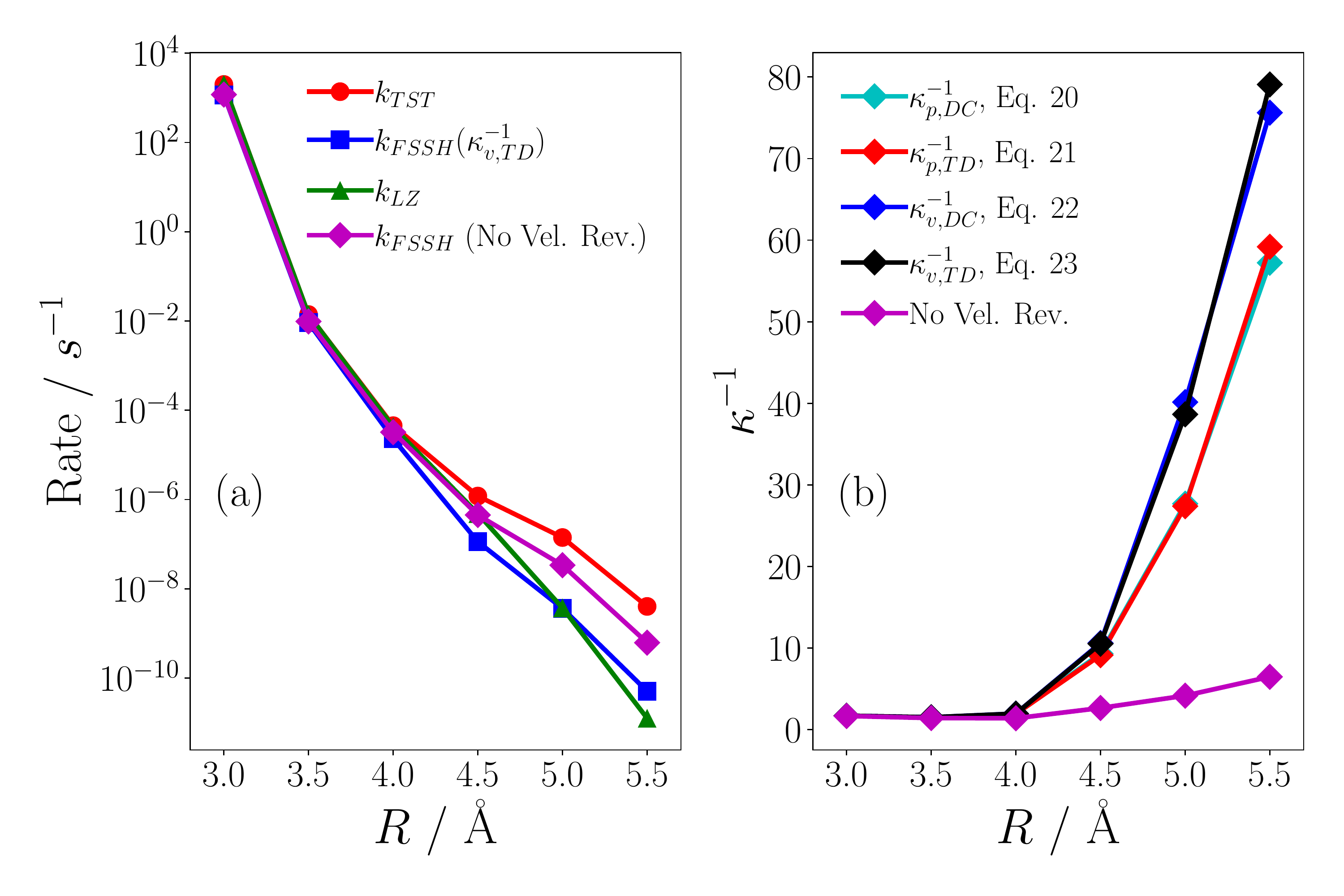}
\caption{\label{Figure:rates_all} (a) The rate of electron transfer (ET) as a function of ion separation distance, from 3~\AA \ to 5.5~\AA, shown here as $k_{FSSH}$. 
The transition state rates as estimated from the barrier heights are shown as $k_{TST}$ and the Landau-Zener corrections to these rates are shown as $k_{LZ}$. 
At small separation distances we find, as expected, that the ET is adiabatic and in nearly full agreement with the transition state rate.
As the separation distance increases the deviation of the observed surface hopping rates from the transition state rate suggests that these systems are much more nonadiabatic in nature.
(b) $\kappa^{-1}$ values using different velocity reversal criteria.
$v$ represents a bare velocity, while $p$ represents a mass-weighted (momentum) velocity.
We see that the expected behavior, namely a sharper reduction in rate at larger ion separation distances, is more correctly anticipated by the velocity based reversal criteria.
Also, there is little difference in using the target diabat or derivative coupling based approaches.}
\end{figure}

The key observable of interest is the rate of crossing from one side of the FES to the other, or the rate of ET between the two stationary ions. 
The expectation is that as the ions are further separated the system will display more diabatic character and the rate of ET between the two ions will decrease.
In Figure \ref{Figure:rates_all}, the TST, Landau-Zener corrected (LZ), and surface hopping rate constants are plotted as a function of ion separation distance.
The LZ rate, $k_{LZ}$ is just the product of the TST rate, $k_{TST}$ and the geometrically summed LZ correction,
\begin{equation}
k_{LZ} = \kappa_{LZ} \cdot k_{TST} = \frac{2 P_{LZ}}{1 + P_{LZ}} k_{TST}
\end{equation}
\noindent where $P_{LZ}$ is defined in Eq.~\ref{eq:k_LZ}.
The results shown in Fig.~\ref{Figure:rates_all}(a) demonstrate that at small ion separation distances, the surface hopping simulation results agree with the TST rate constant, as expected in the adiabatic limit.
As the ion separation distance increases, we see an increase in nonadiabatic behavior, at which point the surface hopping rates are far lower than the TST rates but still in agreement with the Landau-Zener rates. 
In general, this figure broadly endorses the accuracy of the NA-TST approach we describe.

Next, in Fig.~\ref{Figure:rates_all}(b), we plot the ratio of the TST rate to the SH rate, $\frac{k_{TST}}{k_{FSSH}} = \kappa^{-1}$ (\latin{i.e.} the inverse transmission factor) which quantifies how adiabatic the ET event is at a given ion separation distance. 
For these problems, at 5.5~\AA, $\kappa$ can be as small as 1/80, meaning that only 1 out of 80 trajectories crossing a barrier will actually lead to electron transfer.
Interestingly, according to this figure, in order to recover such a small transmission factor, velocity reversal is absolutely essential in the SH simulations, a fact we will now explore in great detail. Without performing velocity reversal, one recovers a rate that is far too large.

\subsection{Velocity reversal criteria}
\begin{figure}[!htb]
\includegraphics[width=16cm,height=10.67cm]{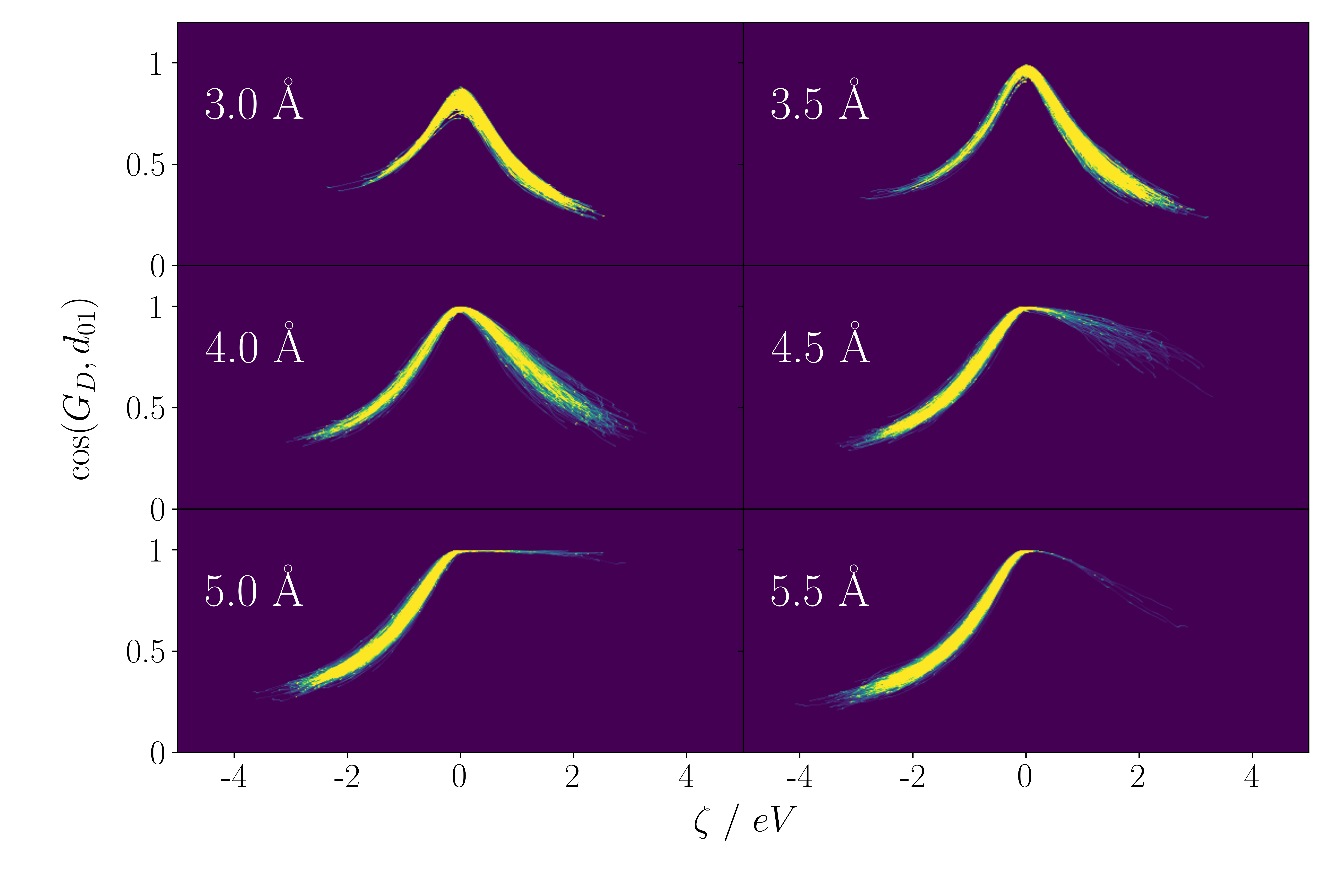}
\caption{\label{Figure:q_vs_cos(gd,d)} Heat maps of the cosine between the difference in diabatic gradients and the derivative coupling, over 500 timesteps from 100~fs for roughly 500 surface hopping trajectories, with ion separation distances from 3~\AA \  to 5.5~\AA.
For all separation distances we see the colinearity between the derivative coupling and the difference in diabatic gradients peak at the barrier region and smoothly decay towards the well representing either ion with the electron.
The asymmetry between left and right well intensity arises from a smaller fraction of total trajectories going to the right in the larger ion separation distance trajectories.
These plots suggest that using velocity reversal criteria based on either the derivative coupling or diabatic gradient difference should be roughly the same, as these quantities point in the same direction in the hopping region, around $\zeta = 0$.}
\end{figure}

\begin{figure}[!htb]
\includegraphics[width=16cm,height=10.67cm]{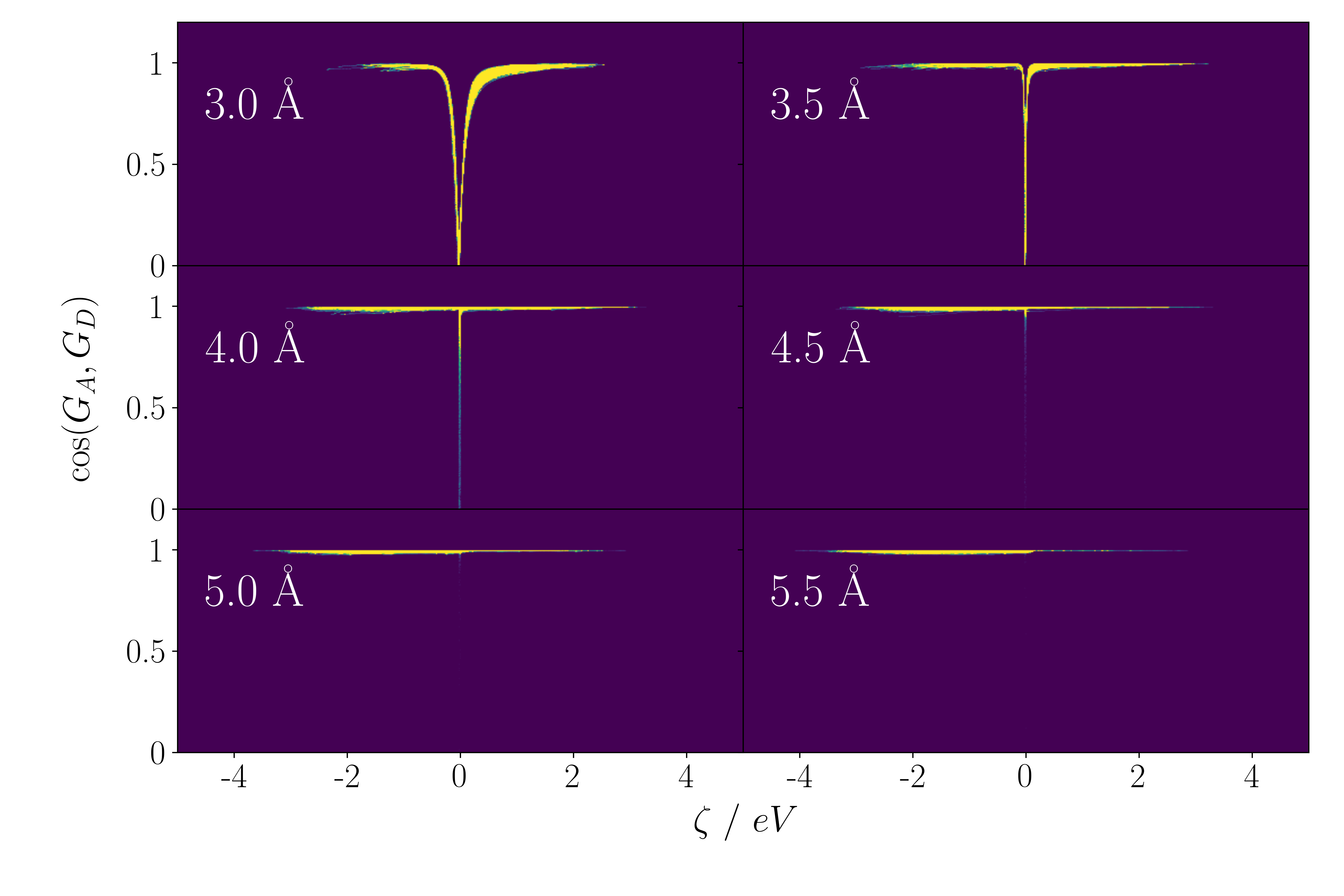}
\caption{\label{Figure:q_vs_cos(ga,gd)} Heat maps of the cosine between the difference in diabatic and adiabatic gradients, over 500 timestepes from 100~fs for roughly 500 surface hopping trajectories, with ion separation distances from 3~\AA \  to 5.5~\AA.
For larger ion separation distances, these two gradient difference are nearly colinear along all regions of the reaction coordinate $\zeta$, but at smaller ion separation distances they exhibit much more orthogonality near the barrier region.
This arises since these separation distances correspond to a more adiabatic transfer event, where the strong coupling in this region necessarily suggests that these two values (the diabatic and adibatic gradient differences) should not be aligned; see Eqs.~\ref{eq:d_and_zeta} and \ref{eq:GA_and_GD}.
Overall, this result implies that using velocity reversal criterion based on the adiabatic gradient difference can give different results than those based on the diabatic gradient difference.}
\end{figure}

A failure to reverse velocities upon a frustrated hop is known in the literature to lead to errors\cite{Muller1997} and often transmission rates through a crossing that are far too large\cite{Schwerdtfeger2014,Jain2015a,Carof2017};
and yet, a simple gedanken experiment demonstrates that one should not always reverse velocities. 
For instance, if two potentials are both driving a trajectory to the right, it would not make sense to implement velocity reversal if the end result were a trajectory moving to the left.

With this background in mind, for a system with one or few degrees of freedom, a common way to determine whether a frustrated hop should undergo velocity reversal is to utilize the technique recommended by Jasper and Truhlar\cite{Jasper2002}.
In this approach, the velocity is reversed along the direction of the derivative coupling between the active and target state (states 0 and 1, respectively), if 
\begin{equation} \label{eq:dc_mo}
(\vec{F}^{a}_{1} \cdot \vec{d}_{01})(\vec{d}_{01} \cdot \vec{p}) < 0, 
\end{equation}
\noindent where $F^{a}_{1}$ is the force on the target adiabatic surface.
Now, Ref.~\citenum{Jasper2002} was based on scattering calculations for a system with a handful of atoms, and it is not clear that this criterion and reversal direction are well-suited or optimal for a system with $10^{3}$ to $10^{4}$ degrees of freedom, even if the barrier crossing might look one dimensional (as we hypothesized above).

To that end, in Fig.~\ref{Figure:rates_all}(b), we have also explored a different reversal criterion based on the diabatic gradient difference. Namely, we reverse if: 
\begin{equation} \label{eq:td_mo}
(\vec{F}^{d}_{u} \cdot \vec{G}_{\textrm{D}})(\vec{G}_{\textrm{D}} \cdot \vec{p}) < 0,
\end{equation}
\noindent Here $F^{d}_{u}$ is the force on the diabatic state with higher energy at the point of the hopping event and $\vec{G}_{\textrm{D}} = \vec{\nabla} \zeta$ (Eq.~\ref{eq:gd_zeta}).
For Eq. \ref{eq:td_mo}, if we choose to reverse, we reverse in the direction $\vec{G}_{\textrm{D}}$.
Moving forward, we will refer to this criteria as the ``target diabat'' criteria.

In addition to the two criteria above, for reasons that will be clear below, we also investigated the analogous expressions whereby we replace the momentum $p$ with the velocity $\vec{v} = \mat{M}^{-1}\vec{p}$:
\begin{align}
(\vec{F}^{a}_{1} \cdot \vec{d}_{01})(\vec{d}_{01} \cdot \vec{v}) < 0 \label{eq:dc_vel} \\
(\vec{F}^{d}_{u} \cdot \vec{G}_{\textrm{D}})(\vec{G}_{\textrm{D}} \cdot \vec{v}) < 0, \label{eq:td_vel}
\end{align}
According to Fig.~\ref{Figure:rates_all}(b), at least for this model system, there is very little difference in using the Truhlar or target diabat criteria.
However, we find that the choice of velocity versus momentum remains important, and that the expected result of a more reduced rate $k_{FSSH}$ as compared to $k_{TST}$ is obtained when using velocity instead of momentum.
To understand why this distinction exists, suppose we wish to follow  a reaction coordinate  based on diabatic energies (\latin{e.g.} $\zeta$) and reverse on the basis of the direction of $\dot\zeta$ when one passes the crossing point and encounters a frustrated hop. According to the chain rule, 
\begin{equation}
   \frac{d\zeta}{dt} = \frac{d\zeta}{d\vec{R}}\frac{d\vec{R}}{dt} = \vec{G}_{\textrm{D}} \cdot \vec{v} 
\end{equation}
and so only if one uses velocities and not momenta (\latin{i.e.} one uses Eq.~\ref{eq:td_vel} instead of Eq.~\ref{eq:dc_mo}) is one guaranteed to reverse one's trajectory.
In our efforts to find a rigorous velocity reversal criterion, we found that the above expression consistently yielded the lowest surface hopping rates.
This realization originally led to our explorations with Eq.~\ref{eq:td_vel}.
Interestingly, although not shown, many other velocity reversal criteria were not effective for this model Hamiltonian (which can behave in a very {\bf not} one-dimensional fashion).

The fact that the two velocity reversal criteria in Eqs.\ \ref{eq:dc_vel} and \ref{eq:td_vel} give such similar results strongly implies that the direction of $\vec{G}_{\textrm{D}}$ and $\vec{d}_{01}$ are nearly colinear, at least at the time and location of a surface hop.
One way to confirm such a hypothesis is to measure the cosine of the angle between $\vec{d}_{01}$ and $\vec{G}_{\textrm{D}}$, which is shown in Fig.~\ref{Figure:q_vs_cos(gd,d)} for different ion separation distances, collated over roughly 500 independent SH trajectories per separation distance each with 500 time steps (so $\sim$ 25000 data points).
From this figure, it is now more evident why the two reversal criteria function nearly identically; in the region around the barrier, $\zeta \approx 0$, the two quantities $\vec{d}_{01}$ and $\vec{G}_{\textrm{D}}$ are nearly colinear.  
This colinearity can be explained by noting that, for a system with two relevant electronic states, a simple diabatic-to-adiabatic transformation shows:
\begin{align}
H &= 
\begin{pmatrix}
\zeta/2 & V \\ 
V & \zeta/2 \\
\end{pmatrix} \\
 \vec{d}_{01} &= \frac{\zeta V}{\Delta E^2}\left(\frac{\vec{\nabla} V}{V} - \frac{\vec{G}_{\textrm{D}}}{\zeta} \right) \label{eq:d_and_zeta}
\end{align}
Thus, for $\zeta$ close to zero, $\vec{d}_{01}$ should be roughly in the direction of $\vec{G}_{\textrm{D}}$.  
From this analysis, one would also conjecture that there there should be no colinearity far away from the crossing for a multi-dimensional problem. 
Indeed, Fig.~\ref{Figure:q_vs_cos(gd,d)} also confirms such a hypothesis: $\vec{d}_{01}$ and $\vec{G}_{\textrm{D}}$ point in very different directions for large $\abs{\zeta}$.

To highlight the fact that this model problem is far from one dimensional in practice, in Fig.~\ref{Figure:q_vs_cos(ga,gd)} we plot the cosine between the diabatic gradient difference $\vec{G}_{\textrm{D}}$ and adiabatic gradient difference
\begin{equation}\label{eq:adia_grad}
\vec{G}_{\textrm{A}} = \vec{\nabla}E_1 - \vec{\nabla}E_0
\end{equation}
\noindent as a function of ion separation distance, again collected over the $\sim$ 25000 data points.  
The data in Fig.~\ref{Figure:q_vs_cos(ga,gd)} is strikingly different from the data in Fig.~\ref{Figure:q_vs_cos(gd,d)}.
In this case, near the barrier region we find that the the adiabatic and diabatic gradient differences are almost entirely orthogonal (note that this difference is also true at larger ion separation distances; the signature is just visually fainter due to a sparsity of data in this region).
This finding can be easily confirmed by expressing the adiabatic gradient as a function of the diabatic energies and couplings:
\begin{align}\label{eq:GA_and_GD}
\vec{G}_{\textrm{A}} = \frac{1}{\zeta \Delta E}\bigl(\zeta \vec{G}_{\textrm{D}} + 4 V \vec{\nabla} V \bigr) 
\end{align}
\noindent Now, for $\zeta$ close to zero, $\vec{d}_{01}$ should be roughly in the direction of $\vec{\nabla}V.$

This finding suggests that velocity reversal criteria methods  using the adiabatic gradient $\vec{G}_{\textrm{A}}$ instead of the  diabatic gradient $\vec{G}_{\textrm{D}}$ (Eq. \ref{eq:td_vel}) or the derivative coupling vector  $\vec{d}_{01}$ (Eq. \ref{eq:dc_vel}) will not lead to equivalent results. Indeed, our cursory surface hopping results (not shown) suggest that using a $\vec{G}_{\textrm{A}}$-based velocity reversal criterion does strongly overestimate the rate in the nonadiabatic limit (relative to the LZ result).

\section{Discussion and Conclusions}\label{sec:discussion}
The above results mark one new step in understanding thermally mediated nonadiabatic transitions in condensed phases using QM/MM methods with explicit solvent.
In this work we have examined the nonadiabatic electron transfer between two fixed ions in acetonitrile. 
As far as rate constants are concerned, we confirmed that, as the separation between the two ions increases, the electronic coupling between the two decreases exponentially and the ET becomes decidedly more nonadiabatic. 
More interestingly, we examined solvent participation ratios and relationships between adiabatic and diabatic quantities as a function of the spatial separation of the two ions.
In particular, as a function of ion separation distance, we constructed heat maps of the relative solvent contributions and calculated the average participation ratio of solvent to hopping events, so that one could physically and atomistically understand the nature of the condensed phase fluctuations that allow for electron transfer.

Looking forward, there are are many next steps for this research, some focused on method development and some focused on applications. 
On the method development side, we note that the above simulations did not include decoherence effects between different adiabatic surfaces.
While the present model problem may be minimally affected by wavepacket coherence, there are  nonadiabatic problems in the condensed phase where decoherence cannot be easily ignored\cite{landry:2011:marcus_fssh}.
Given that most models of decoherence have been tested most strenuously on small  systems (where exact bechnmarks are possible), it likely that new physics  and improvements to decoherence will emerge by including such effects within NA-TST calculations in  the future. 

Another computational issue relates to velocity reversal. 
The problem of how best to handle velocity reversal may not have a uniform, one-size-fits-all solution, and will certainly need to be tested on a variety of condensed phase systems to obtain a better understanding of the relative strengths and restrictions of various methods. 
In particular, above we argued that one could improve upon Eq.~\ref{eq:dc_mo} simply by replacing the momentum with the velocity (Eq.~\ref{eq:dc_vel}). 
Nevertheless, we were still able to construct two different velocity reversal approaches that behaved just about as well as each other (Eqs.\ \ref{eq:dc_vel} and \ref{eq:td_vel}). 
Will one of these criteria perform better in general? Obviously, one would prefer Eq.~\ref{eq:dc_vel} to Eq.~\ref{eq:td_vel} if possible (so that one uses a definition of diabatic states as minimally as possible).  
Are there still other, better approaches out there?

Turning to applications, the most obvious next step is to run \latin{ab initio} NA-TST simulations with the Q-Chem/INAQS interface. There is no computational bottleneck and there are a plethora of problems that one would like to tackle with NA-TST \cite{Carpenter2006}.
Finally, another very promising avenue for future research will be to extend the current simulations to condensed phase systems with a continuum of electronic states, such as the case of a metal electrode in interface with a liquid solvent, as is often seen in various electrochemical applications.
One application of great interest is that of the Volmer-Heyrovsk\'{y} reaction\cite{Heyrovsky1927,Erdey-Gruz1930}, whereby a surface adsorbed hydrogen recombines with a solvated proton to form molecular hydrogen. 
Inclusion of an explicit metal surface in the form of a tight binding model would allow for a system such as this one to be simulated in a form analogous to the simulations presented here.

The above examples of future thrusts are just a small selection of a myriad number of ways the QM/MM framework here within could be extended to study nonadiabatic phenomena in condensed phases, allowing for a very powerful new understanding of these processes at a molecular level. 

\begin{acknowledgement}
This work was supported by the U.S. Air Force Office of Scientific Research (USAFOSR) (under Grant Nos.\ FA9550-18-1-0497 and FA9550-18-1-0420). 
We thank the DoD High Performance Computing Modernization Program for computer time.
\end{acknowledgement}

\bibliography{bibliography, bibliography_new}

\begin{mcitethebibliography}{78}
\bibitem[Not()]{Note-1}
Only at $R=4.0\,\textrm{\AA}$ do we have sufficient trajectories that are
  reactive (272) or nonreactive (194) that we can make a statistically
  meaningful comparison.\relax
\mciteBstWouldAddEndPunctfalse
\mciteSetBstMidEndSepPunct{\mcitedefaultmidpunct}
{}{\mcitedefaultseppunct}\relax
\EndOfBibitem
\end{mcitethebibliography}
\end{document}